\documentclass{emulateapj}

\usepackage{lineno}
\usepackage{graphicx}
\usepackage{multirow}
\usepackage{natbib}
\usepackage{color}
\usepackage{xspace}
\usepackage{hyperref}
\usepackage[normalem]{ulem}

\makeatletter

\newcommand{\Rmnum}[1]{\expandafter\@slowromancap\romannumeral #1@}
\makeatother
\newcommand{\fermi}{\textit{Fermi}-LAT\xspace}
\newcommand{\hess}{H.E.S.S.\xspace}
\newcommand{\rxj}{RX~J1713.7$-$3946\xspace}
\newcommand{\rcw}{RCW~86\xspace}
\newcommand{\vela} {RX~J0852.0$-$4622\xspace}
\newcommand{\hjdt} {HESS~J1731$-$347\xspace}
\newcommand{\hjdv} {HESS~J1729$-$345\xspace}
\newcommand{\sn} {SN~1006\xspace}

\begin{document}

\title{Detection of two TeV shell-type remnants at GeV \\ energies with \textit{FERMI} LAT: \hjdt and \sn}

\author{
B.~Condon\altaffilmark{1,2}, 
M.~Lemoine-Goumard\altaffilmark{1}
F.~Acero\altaffilmark{3}, 
H.~Katagiri\altaffilmark{4}, 
}
\altaffiltext{1}{Centre d'\'Etudes Nucl\'eaires de Bordeaux Gradignan, IN2P3/CNRS, Universit\'e Bordeaux 1, BP120, F-33175 Gradignan Cedex, France}
\altaffiltext{2}{email: condon@cenbg.in2p3.fr}
\altaffiltext{3}{Laboratoire AIM, CEA-IRFU/CNRS/Universit\'e Paris Diderot, Service d'Astrophysique, CEA Saclay, F-91191 Gif sur Yvette, France}
\altaffiltext{4}{College of Science, Ibaraki University, 2-1-1, Bunkyo, Mito 310-8512, Japan}


\begin{abstract}
We report the first high-significance GeV gamma-ray detections of supernova remnants \hjdt and \sn, both of which have been previously detected by imaging atmospheric Cherenkov Telescopes above 1~TeV. Using 8 years of \fermi Pass 8 data at energies between 1~GeV and 2~TeV, we detect emission at the position of \hjdt with a significance of $\sim 5\sigma$ and a spectral index of $\Gamma = 1.66 \pm 0.16_{\rm stat} \pm 0.12_{\rm syst}$. The hardness of the index and the good connection with the TeV spectrum of \hjdt support an association between the two sources. We also confirm the detection of \sn at $\sim 6\sigma$ with a spectral index of $\Gamma = 1.79 \pm 0.17_{\rm stat} \pm 0.27_{\rm syst}$. The northeast (NE) and southwest (SW) limbs of \sn were also fit separately, resulting in the detection of the NE region ($\Gamma = 1.47 \pm 0.26_{\rm stat}$) and the non-detection of the SW region. The significance of different spectral components for the two limbs is $3.6\sigma$, providing first indications of an asymmetry in the GeV $\gamma$-ray emission.\\
\end{abstract}

\maketitle


\section{Introduction}\label{sec:introduction}
\setcounter{footnote}{0}

With the detection of X-ray synchrotron emission coming from the shock wave of \sn, \cite{Koyama1995} proved that supernova remnants (SNR) were able to accelerate electrons (and probably protons as well) up to a few hundred TeV. Since then, SNRs radiating synchrotron X-rays have been closely studied as potential cosmic-ray (CR) accelerators. Another evidence of particle acceleration is the presence of $\gamma$ rays induced by the inverse Compton scattering of relativistic electrons on ambient photon fields (leptonic scenario) or the decay of neutral pions produced by proton-proton interactions (hadronic scenario). The current generation of imaging atmospheric Cherenkov telescopes such as \hess, MAGIC and VERITAS, has measured very-high energy $\gamma$-rays ($0.3-10$~TeV) coming from several SNRs detected in X-rays, confirming the presence of ultra-relativistic particles.

In the early 2000's, the CANGAROO collaboration reported the first TeV detection of \rxj\ \citep{Muraishi2000, Enomoto2002} and \vela \citep{Katagiri2005}. The \hess experiment later confirmed those two detections \citep{Aharonian2004, Aharonian2005} and revealed that $\gamma$-ray emission was also coming from \rcw \citep{Aharonian2009}, \sn \citep{Acero2010} and \hjdt \citep{Aharonian2008}. The latter was discovered during the first \hess Galactic Plane Survey \citep{HGPS2005, HGPS2006} and had no known counterpart. Their large structures allow detection of $\gamma$ rays coming from the shock wave itself. They constitute a type of SNR of great interest: TeV shell-type remnants. Their common characteristics are a young age (less than a few thousand years old), a large angular size ($R_{\rm SNR} > 0\fdg25$) and TeV emission highly correlated with X-ray synchrotron emission.

However, the TeV spectrum alone is not sufficient to distinguish between a leptonic and hadronic scenario. We need to study all these SNRs at GeV energies to characterize the shape of the spectrum in a wide energy range and be able to identify the mechanism of $\gamma$-ray production. Among the five TeV shell-type remnants, three have already been detected and carefully studied at GeV energies: \vela \citep{Tanaka2011}, \rxj\ \citep{Abdo2011, Federici2015} and \rcw \citep{Ajello2016}. However, \hjdt and \sn remained undetected, though a recent work reported first evidence of $\gamma$-ray emission coming from the latter with $\sim 4\sigma$ significance \citep{Xing2016}.

In this paper we report on the first detection of \hjdt and we confirm the detection of \sn at GeV energies with the \textit{Fermi} Large Area Telescope (LAT).

\section{\fermi and data reduction}
\label{sec:lat}

The Large Area Telescope on board the \textit{Fermi Gamma-Ray Space Telescope} (\textit{Fermi}) mission is a $\gamma$-ray telescope that detects photons by conversion into $e^+e^-$ pairs in the energy range from 20 MeV to greater than 300 GeV. A full description of the instrument can be found in \cite{Atwood2009} and the in-flight performance in \cite{Ackermann2012}. 

Both sources were studied in the same conditions (identical data selection and analysis methods). For this work, we analyzed 8 years (August 4$^{\rm th}$ 2008 $-$ August 28$^{\rm th}$ 2016) of LAT Pass~8 data \citep{Atwood2013}, selecting photons with a reconstructed energy between 1 GeV and 2 TeV in order to lower the pollution from the Galactic diffuse background, which is crucial for \hjdt. We also rejected events with zenith angle greater than $100^\circ$ to reduce the contamination from the Earth limb and applied the recommended cuts {\tt (DATA\_QUAL)==1 \&\& (LAT\_CONFIG)==1} to assure good quality data. Moreover, time intervals when the \textit{Fermi} spacecraft was within the Southern Atlantic Anomaly were excluded, as well as those when the Sun or the Moon was crossing the region of interest. The version {\tt v11r05p01} of the \textit{Fermi} Science Tools was used with the {\tt P8R2\_SOURCE\_V6} Instrument Response Function (IRFs).

Two different tools were used: {\tt pointlike} for the spatial analysis and {\tt gtlike} for the spectral analysis. The first one is a code developed by members of the LAT collaboration and optimized for the characterization of extended sources \citep{Kerr2011}. The second one is implemented in the \textit{Fermi} Science Tools and is the official tool for \fermi analysis. Both are based on a maximum likelihood method \citep{Mattox1996}. 

\section{Analysis Method }
\label{sc:analysis_method}

The analysis was performed in a circular region of $7^\circ$ around each source of interest (SOI), \hjdt and \sn. The initial models included sources from the \fermi Third Source Catalog \cite[][hereafter 3FGL]{3FGL} within $15^\circ$, as well as the Galactic diffuse background ({\tt gll\_iem\_v06.fits}) and extragalactic isotropic background ({\tt iso\_P8R2\_SOURCE\_V6\_v06.txt}). These background models are available on the FSSC website\footnote{https://fermi.gsfc.nasa.gov/ssc/data/access/lat/} and are described in details in \cite{Acero2016a}. The spectral parameters of sources located within $5^\circ$ as well as those of the Galactic (normalization and index) and isotropic (normalization only) diffuse emissions were set free.

\begin{figure}[ht]
  \centering
  \includegraphics[width=\columnwidth]{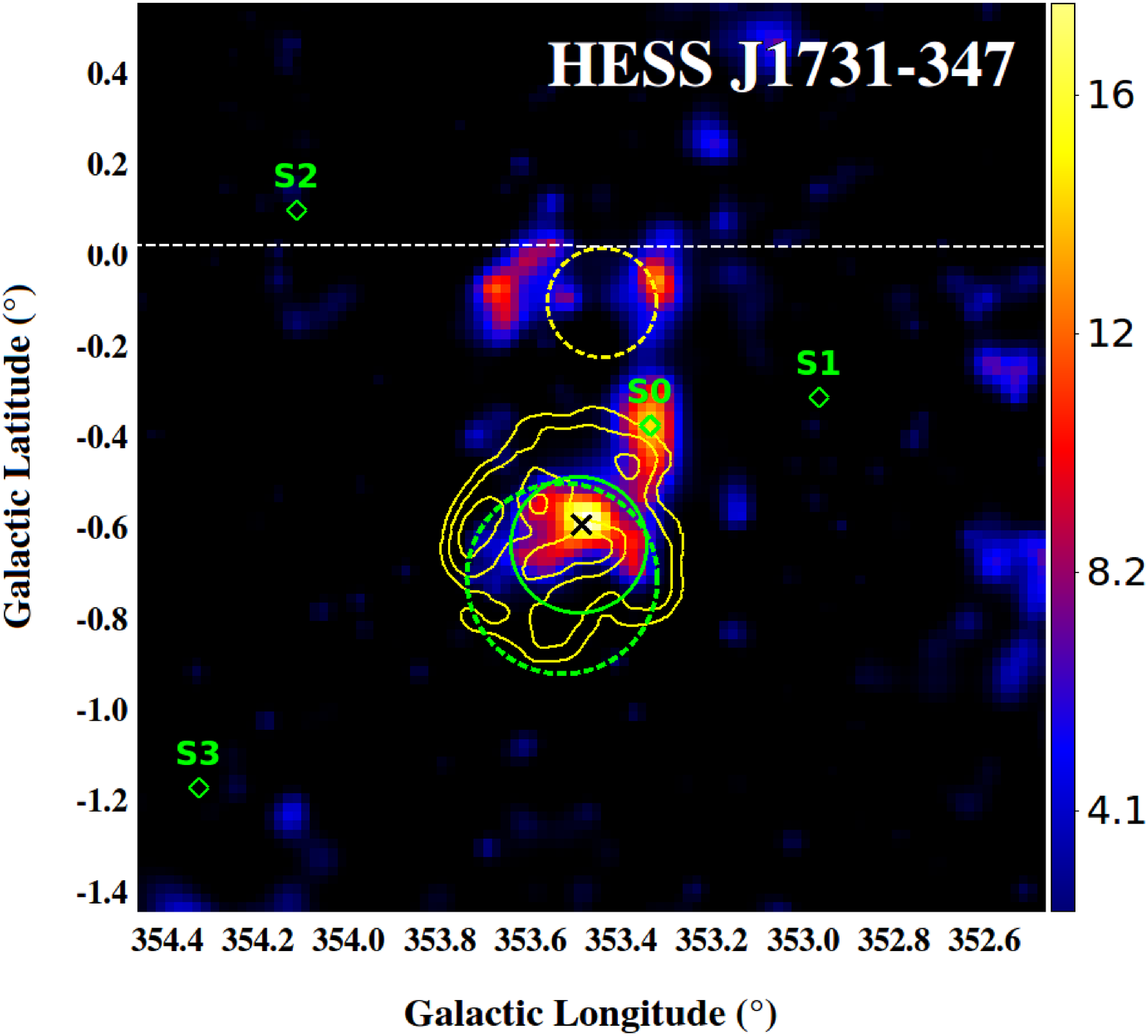}
  \includegraphics[width=\columnwidth]{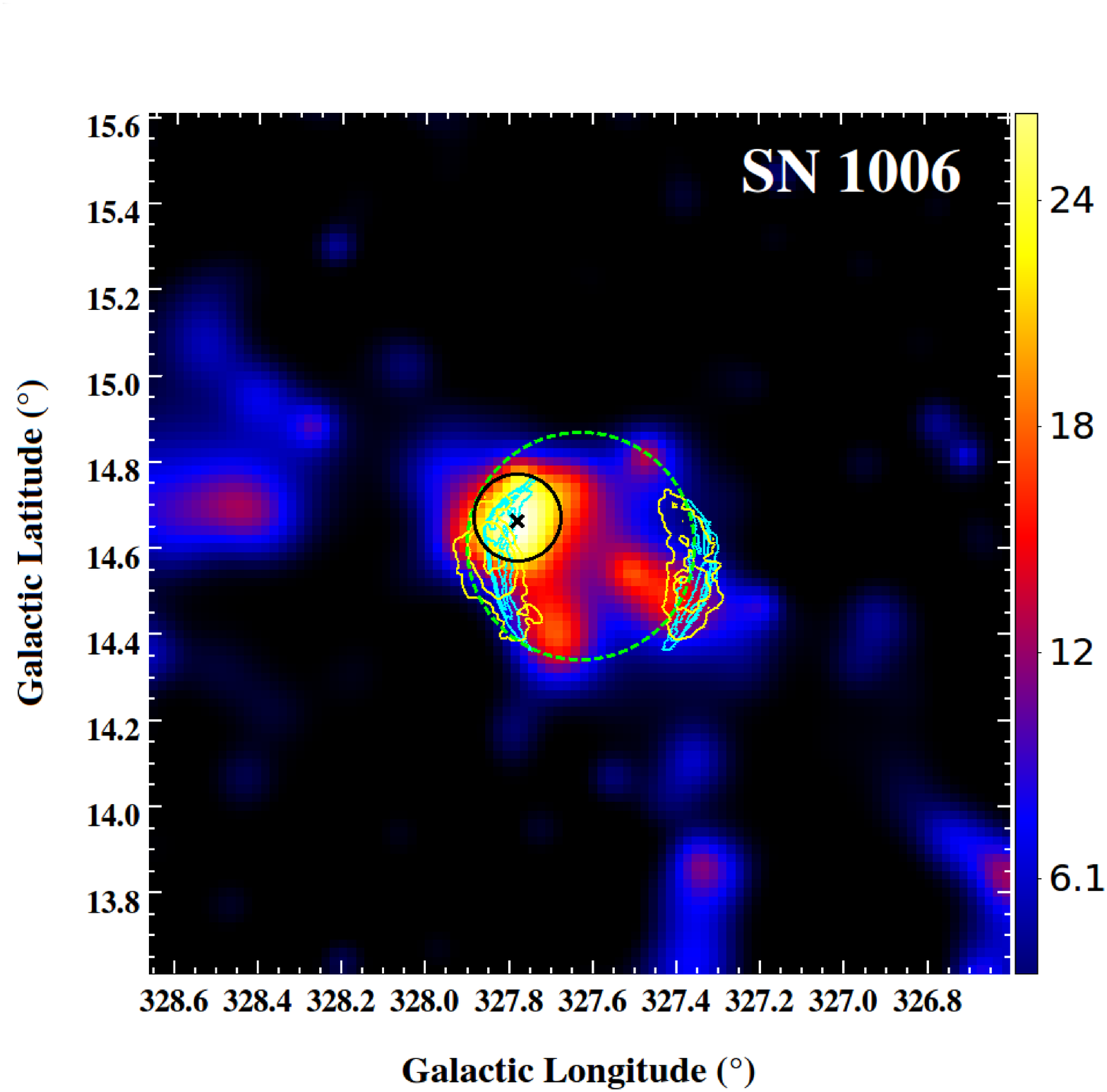}
  \caption{TS maps (1 GeV $-$ 2 TeV) in Galactic coordinates showing a $2^\circ \times 2^\circ$ region centered on the position of \hjdt (top) and \sn (bottom). In the upper panel, green diamonds correspond to additional background sources while the black cross and green circle correspond to the best-fit position (for $E>1$~GeV) of \hjdt as a point source and a uniform disk respectively. The best uniform disk above 10 GeV is represented by the green dashed circle. The TeV morphology of the shell is represented by the yellow contours and the yellow dashed circle shows the position of \hjdv. In the lower panel, the best-fit position of \sn as a point source and a uniform disk are represented by the black cross and circle, respectively. The green dashed circle shows the result obtained with an initial disk matching the TeV shape. Yellow and cyan contours correspond to \hess and XMM-Newton, respectively. No background sources were found in the close vicinity of \sn. \label{fig:tsmap}}
\end{figure}

To search for GeV counterparts to \hjdt and \sn, we first computed a $10^\circ \times 10^\circ$ Test Statistic (TS) map, which is obtained by evaluating the TS value \cite[as described in ][]{Mattox1996} of a point source with a fixed index of 2.0 in each pixel of the map. Gamma-ray excesses with a TS higher than 25 were added to the model as new point sources. As shown in Figure~\ref{fig:tsmap}, the TS maps revealed $\gamma$-ray excesses at the position of \hjdt and \sn. Therefore, we performed a spatial analysis with {\tt pointlike} to characterize the shape of these excesses. The two sources were modeled as point sources and uniform disks. For these two geometrical models, the spatial parameters were adjusted while fitting the spectrum (normalization and index) simultaneously. In the case of an extended source (e.g., a uniform disk), the significance of the extension was quantified by computing the TS$_{\rm ext}$ value \citep{Lande2012} with {\tt pointlike}. To compare the morphology detected in the GeV range with the one observed at higher energy by imaging atmospheric Cherenkov Telescope, we also fit the LAT data with templates derived from \hess excess maps. The results of the spatial analysis performed with {\tt pointlike} were cross-checked with {\tt gtlike}. For both sources, we performed binned ({\tt pointlike}) and unbinned ({\tt gtlike}) likelihood analyses in a region centered on the position ($l$, $b$) = ($353\fdg49$, $-0\fdg62$) for \hjdt and ($l$, $b$) = ($327\fdg63$, $+14\fdg61$) for \sn.

After having characterized the morphology, we studied the spectra of the two SOIs, both modeled by a power law. Their spectral parameters were fit between 1 GeV and 2 TeV along with those of background sources located within $5^\circ$ around them. To compute the spectral energy distribution (SED), the 1 GeV $-$ 2 TeV energy range was divided into 4 logarithmically spaced bins and we performed a fit of each of them with the spectral index of the SOI being fixed at 2.0. A spectral point was kept for a TS $>$ 3 and at least three photons associated to the SOI in the considered bin. Otherwise, an upper limit at 95\% confidence level (C.L.) was derived using a Bayesian method \citep{Helene1983}.

For both the global fit and the SED, three types of systematic errors were taken into account: the imperfect modeling of the Galactic diffuse emission, uncertainties in the effective area calibration and uncertainties in the source shape. The first one was estimated by using alternative Interstellar Emission Models (IEMs) as described in \cite{SNRCAT}. For the second one, we used modified IRFs whose effective areas bracket the nominal ones \citep{Ackermann2012}. The last one was estimated by taking into account the variation of the results obtained when modeling the source with different spatial shapes (a point source, a uniform disk and the \hess template). The total systematic error was obtained by computing the quadratic sum of the three different errors.

\section{\hjdt}
\subsection{Results}
\label{sc:hjdt_results}

\hjdt is located in a complex region, close to the Galactic plane and the Galactic center. In the $10^\circ \times 10^\circ$ region, seventeen sources were detected with a TS higher than 25 and thus added to the model. To avoid any contamination, we also took into account an excess (called 'S0') with a TS of only 22 because of its position, very close to the SOI. The upper panel of Figure~\ref{fig:tsmap} shows the $2^\circ \times 2^\circ$ TS map centered on the position of \hjdt. In this case, S0 was not included in the model in order to show its corresponding $\gamma$-ray emission. The TeV emission detected by \hess in this region is associated with two different sources: the shell-type SNR \hjdt and an unidentified source called \hjdv \citep{Acero2011}. Since we are studying the GeV counterpart of \hjdt, the template derived from the \hess excess map was restricted to the shell region, the TeV emission towards \hjdv being removed.

\begin{figure}[ht]
  \centering
  \includegraphics[width=\columnwidth]{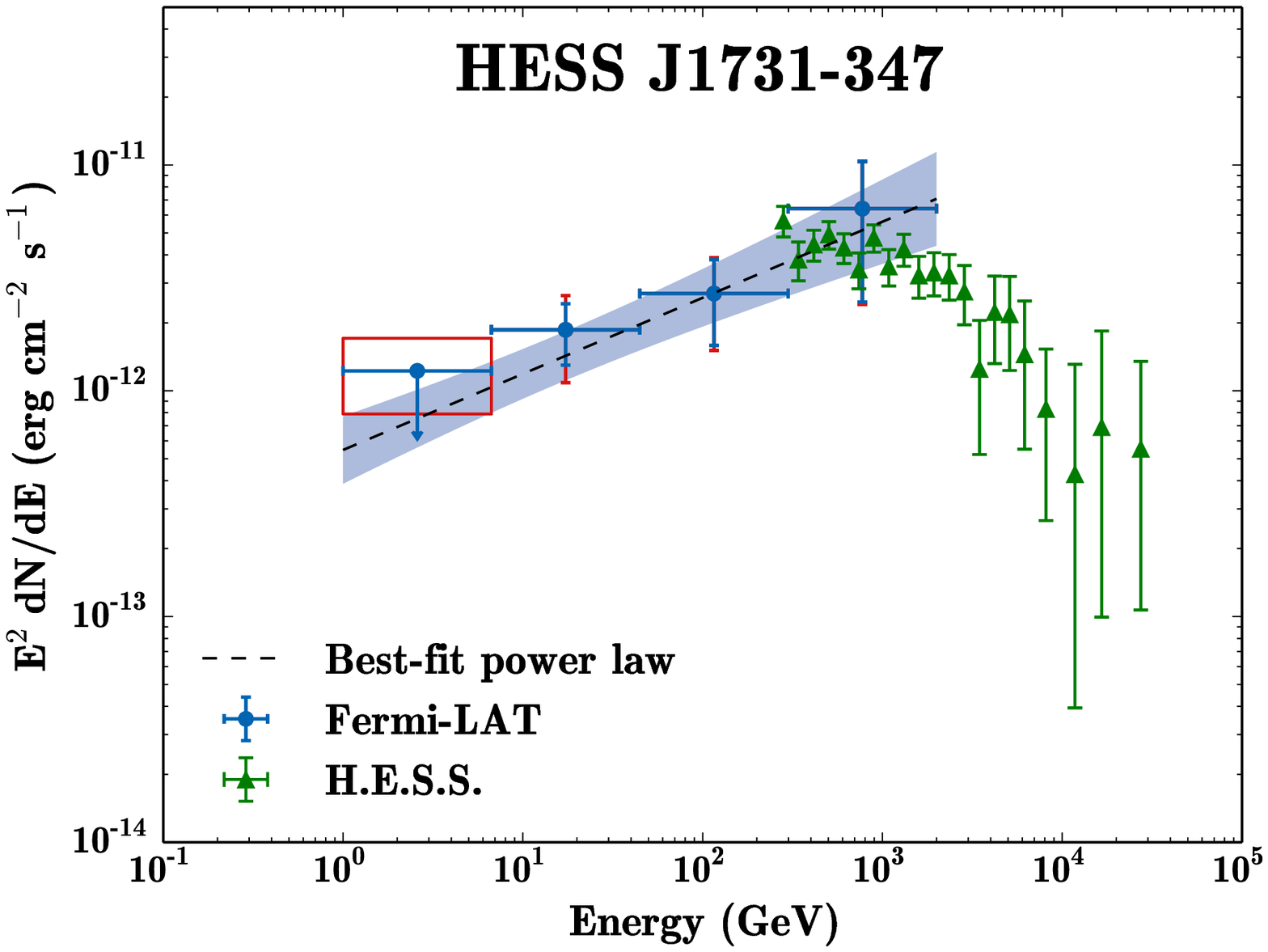}
  \includegraphics[width=\columnwidth]{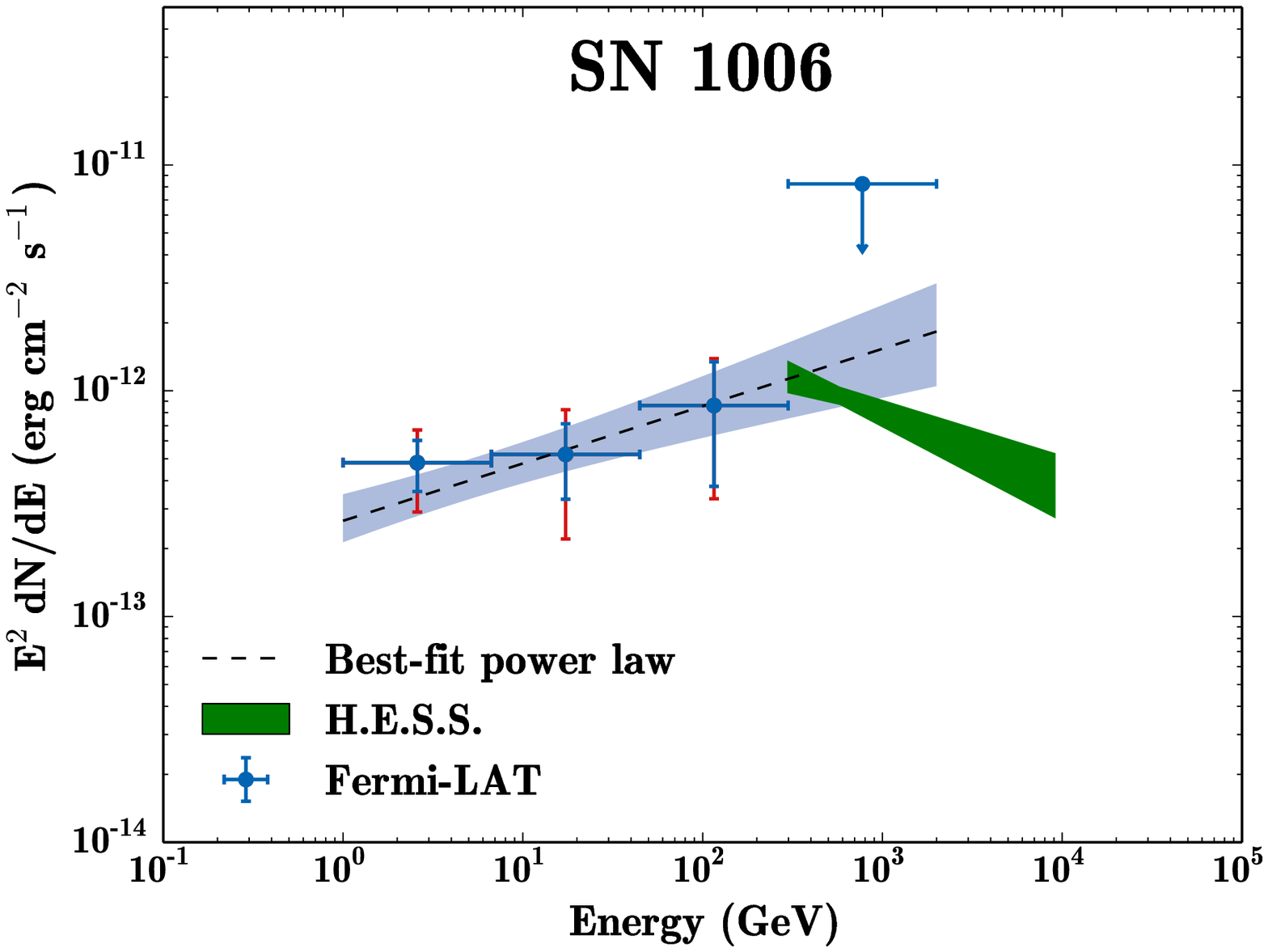}
  \caption{Spectral Energy Distribution for \hjdt (top) and \sn (bottom). In both panel, blue points correspond to the \fermi data and blue shaded areas represent the 68\% confidence band of the \fermi spectral fit (statistical errors). Red error bars are the quadratic sum of statistical and systematic errors. In the upper panel, the variation of the upper limit with alternative IEMs is represented by the red rectangle that surrounds it and the green triangles represent the \hess data points. In the lower panel, the green shaded area corresponds to the power-law fit of the \hess data, taking into account statistical errors. \label{fig:sed}}
\end{figure}

The $\gamma$-ray excess detected by \fermi was first fit as a point source and then as a uniform disk. Their positions are represented in Figure~\ref{fig:tsmap} (top) by the black cross and the green circle, respectively, and the precise value of each parameter can be found in Table~\ref{tab:results}. In the case of a uniform disk, we obtained a radius of $0\fdg15 \pm 0\fdg04$ and a TS$_{\rm ext}$ value of 3.0, indicating that the source is not detected as extended. If the search for an extension of the source is performed above 10 GeV, the analysis yield a TS$_{\rm ext}$ of 7.2 and a larger radius ($0\fdg21 \pm 0\fdg03$, see upper panel of Figure~\ref{fig:tsmap}). Although it is still not significant, it indicates that the source is probably extended and that we are only limited by the low statistics. This pattern has been observed with other LAT sources that eventually proved to be extended after more years of observations. This has been the case of RCW 86, first detected as a point source in \cite{Yuan2014} and then as an extended source in \cite{Ajello2016}.

After having determined the best spatial parameters of the geometrical models, we performed unbinned likelihood analysis with {\tt gtlike} for various morphologies. The fit between 1 GeV and 2 TeV yields a TS value of 25.2, 31.5 and 25.1 for the point source, the disk and the \hess template, respectively. The $\gamma$-ray excess found by \fermi at the position of \hjdt is therefore significant, whatever the assumed shape. Using alternative diffuse emission models also resulted in TS $>$ 25. Regarding the spectral index, the fit yields $\Gamma = 1.87$, $\Gamma = 1.71$ and $\Gamma = 1.66$ respectively, with an error of $\pm 0.16_{\rm stat} \pm 0.12_{\rm syst}$. Although consistent within errors, the spectral index seems to harden when the spatial model gets larger. If real, this could be due to the presence of high-energy photons located in the shell region that are taken into account by the TeV template but not in the hypothesis of a point source. Spectral results are summarized in Table~\ref{tab:results}.

As the GeV emission is at this time consistent with all three of the models tested, we cannot use the morphology to say whether the GeV emission and \hjdt arise from a common source. However, we could explore the question further by deriving an SED using the \hjdt template, under the assumption that the two sources are connected. By doing so, we were able to compare directly the GeV spectral points with the TeV data. The hypothesis that the GeV emission is correctly described by the \hess template can be considered since the likelihood was not significantly degraded when using the \hess template in comparison to the point source or disk hypothesis. As shown in the upper panel of Figure~\ref{fig:sed}, the LAT data (blue points) match very well the TeV data (green triangles), supporting our assumption that the $\gamma$-ray excess detected by the LAT is the GeV counterpart of \hjdt.

We also studied the second TeV source found by \hess in this region, \hjdv. We added a gaussian source in the model \citep[R.A. = $262\fdg39$, Dec. = $-34\fdg54$, $\sigma =0.14^\circ$, ][]{Acero2011} and performed a new fit. As it resulted in a non-detection (TS $<$ 3), we derived a 95\% C.L. upper limit ($1.30 \times 10^{-12}$ erg cm$^{-2}$ s$^{-1}$ between 1 GeV and 2 TeV). This upper limit is a factor of two above the TeV data and cannot constrain the origin of the $\gamma$-ray emission of \hjdv. However, it confirmed a turnover between the \fermi and \hess data.

Finally, previous works brought our attention to the $\gamma$-ray excess called S0. Observations conducted at 1.4 GHz with ATCA and Parkes \citep{SGPS2005} have revealed the presence of an \ion{H}{2} cloud at the position of S0 \cite[see also Figure~5 in ][]{Acero2011}. In addition to the detection of CO lines next to \hjdt, a recent paper investigated the possibility that CRs were diffusing from the SNR forward shock to nearby molecular clouds and made predictions regarding the $\gamma$-ray emission of an hypothetical cloud called MC--core \citep[see Figure~3 in ][]{Cui2016} located at a very close position to S0. For those reasons, we studied the spectrum of S0 in the GeV band and found a spectral index of $\Gamma = 2.50 \pm 0.05_{\rm stat}$ and an energy flux of $(4.90 \pm 1.28_{\rm stat}) \times 10^{-12}$ erg cm$^{-2}$ s$^{-1}$. If S0 corresponds to MC--core, this result does not support \cite{Cui2016}'s models which predicted a hard index at GeV energies. Since S0 could be another background source and not related to MC--core, we added another point source, at the same position. In this situation, no significant signal was found in addition to S0 and the derived upper limit ($9.31 \times 10^{13}$ erg cm$^{-2}$ s$^{-1}$ between 1 GeV and 2 TeV) could not constrain the predicted models.

\begin{table*}[ht]
  \caption{Summary of the results obtained between 1 GeV and 2 TeV for \hjdt and \sn. \label{tab:results}}
  \centering
  \resizebox{2.1\columnwidth}{!}{%
  \begin{tabular}{c|lccccccc}
    \hline
    \hline
    Source & Spatial Model & R.A. ($^\circ$) & Dec. ($^\circ$) & Radius ($^\circ$) & Spectral Index & Energy Flux (erg cm$^{-2}$ s$^{-1}$) & TS & $N_{\rm dof}$\rule[-4pt]{0pt}{12pt}\\
    \hline
    \hline
    \multirow{3}{*}{\hjdt} & Point Source & $262.92 \pm 0.02$ & $-34.77 \pm 0.02$ & $-$             & $1.87 \pm 0.12$ & $(0.78 \pm 0.31) \times 10^{-11}$ & 25.2 & 4\rule[-3pt]{0pt}{12pt}\\
                           & Disk         & $262.97 \pm 0.03$ & $-34.79 \pm 0.02$ & $0.15 \pm 0.04$ & $1.71 \pm 0.17$ & $(1.61 \pm 0.50) \times 10^{-11}$ & 31.5 & 5\rule[-3pt]{0pt}{12pt}\\
                           & \hess        & $-$               & $-$               & $-$             & $1.66 \pm 0.16$ & $(1.91 \pm 0.63) \times 10^{-11}$ & 25.1 & 2\rule[-3pt]{0pt}{12pt}\\
    \hline
    \multirow{5}{*}{\sn} & Point Source & $225.88 \pm 0.04$ & $-41.75 \pm 0.02$ & $-$             & $1.55 \pm 0.10$ & $(5.56 \pm 2.57) \times 10^{-12}$ & 28.6 & 4\rule[-3pt]{0pt}{12pt}\\
                         & Disk         & $225.90 \pm 0.04$ & $-41.74 \pm 0.03$ & $0.10 \pm 0.04$ & $1.57 \pm 0.11$ & $(5.82 \pm 2.60) \times 10^{-12}$ & 31.0 & 5\rule[-3pt]{0pt}{12pt}\\
                         & \hess        & $-$               & $-$               & $-$             & $1.79 \pm 0.17$ & $(6.99 \pm 2.03) \times 10^{-12}$ & 34.4 & 2\rule[-3pt]{0pt}{12pt}\\
                         & \hess (NE)   & $-$               & $-$               & $-$             & $1.47 \pm 0.26$ & $(6.14 \pm 2.53) \times 10^{-12}$ & 28.3 & 2\rule[-3pt]{0pt}{12pt}\\
                         & \hess (SW)   & $-$               & $-$               & $-$             & $2.60 \pm 0.80$ & $(0.88 \pm 0.24) \times 10^{-12}$ & 12.9 & 2\rule[-3pt]{0pt}{12pt}\\
    \hline
  \end{tabular}
  }
  \tablecomments{Parameters of the geometrical models were adjusted with {\tt pointlike} while TS values and spectral indices were obtained using {\tt gtlike}. Only statistical errors are shown here.}
\end{table*}

\subsection{Discussion}

The analysis presented in the previous section shows that a significant $\gamma$-ray excess is found at the position of \hjdt but its association with the SNR must be discussed.

The hard index obtained in the GeV band ($\Gamma \sim 1.66$) and the good connection with the TeV data strongly support the association between the TeV shell and the GeV source. This spectral shape is what we would expect for a young shell-type remnant, as was found for \rxj, \vela or \rcw. This result alone allows us to rule out several alternative explanations for the origin of this $\gamma$-ray excess. Nevertheless, we considered other scenarios and tried to explain why they are less likely than our interpretation.

According to the TS map computed above 1 GeV, the emission seems to come from within the shell, where the TeV flux is the lowest, and the spatial analysis revealed that it is not significantly extended. However, the low statistics ($\sim 150$ photons) could explain such an unexpected shape and with a few years of additional data an extended shape might be observed. Besides, the likelihood did not rule out a large extended source (e.g., a uniform disk with a radius of $0\fdg25$ or the \hess template) in comparison to the point source hypothesis. 

In X-rays, \hjdt was only partially covered  \citep{Tian2008, Tian2010} until a recent study of the SNR with \textit{XMM-Newton} based on observations of the whole remnant \citep{Doroshenko2017}. The authors reported a shape and a size compatible with \hess results but noted a suppression of the X-ray emission towards the Galactic plane which could be explained by a lower velocity of the shock due to its interaction with a molecular cloud. Since our analysis was already completed at that time, we did not compare the LAT data and the new X-ray morphology. In addition to the shell, a compact source named XMMU~J173203.3$-$344518 is detected in X-rays at the geometrical center of the SNR \citep{Tian2010}. It lies at $\sim 0\fdg08$ from the LAT peak and is thought to be the central compact object of \hjdt \citep{Klochkov2015}. But it is unlikely that this X-ray source and the $\gamma$-ray emission detected by the LAT are related because of this spatial shift which is significant by $\sim 4\sigma$ according to the spatial error from {\tt pointlike}. Besides, \cite{Halpern2010} studied this X-ray source as a candidate magnetar, a class of objects not known to be GeV or TeV emitters. In addition, this point source is similar (lack of infrared and optical counterparts, no pulsation, thermal spectrum modeled as blackbody emission with temperature of the order of $0.2-0.5$~keV) to the CCOs found in \rxj \citep{CassamChenai2004} and \vela \citep{Kargaltsev2002}, where the $\gamma$-ray emission stems from the shell and not from the central source. Concerning pulsation, no search could be performed at GeV energies as no ephemerides are available  \citep[no pulsation were found in X-rays by][]{Halpern2010} and the small number of photons makes a blind search impossible.

We also considered the hypothesis that the LAT signal was due to molecular clouds but such clouds have not been found at that position. The observations of an \ion{H}{2} region \citep{Tian2010}, as well as spectral lines from CO \citep{Acero2011} and CS \citep{Maxted2015}, were reported near the position of the background source S0 but nothing within the shell.

Finally, an extragalactic origin can be reasonably ruled out since no active galaxy nuclei (AGN) are known at this position. Based on the Third Catalog of Hard \fermi sources  \citep{3FHL}, the surface density of blazars for a photon flux of $5 \times 10^{-10}$ ph cm$^{-2}$ s$^{-1}$ in this energy range is $\sim 7 \times 10^{-3}$ deg$^{-2}$, corresponding to a rather low probability ($1.4 \times 10^{-3}$) to get such a source on top of \hjdt assuming a $0.5^\circ$ diameter. We also investigated a potential variability of the signal on a yearly basis but did not find any obvious change in the flux.

The weakness of the signal detected prevents us from strong conclusions about this GeV source but we argue that its association with the shell-type SNR \hjdt is the most probable scenario.

\section{\sn}
\subsection{Results \label{sc:sn_results}}

The procedure described in Section~\ref{sc:analysis_method} was followed to study \sn. We first looked for new background sources in a 10$^\circ$ square region and found three significant excesses at some distance from the SOI. As shown in the lower panel of Figure~\ref{fig:tsmap}, the TS map revealed a bright spot of $\gamma$-ray emission at the top of the northeast (NE) limb of \sn, along with some diffuse emission. Using {\tt pointlike} for the spatial analysis, we tested the point source and the uniform disk hypotheses and determined the best parameter values (see Table~\ref{tab:results}). Assuming a uniform disk initially centered on the pixel of highest TS value with $R_{\rm init} = 0\fdg01$, we measured an extension of $0\fdg10 \pm 0\fdg04$ with TS$_{\rm ext} = 1.8$ : \sn is therefore not detected by the LAT as an extended source. The positions of the point source and the disk are represented in Figure~\ref{fig:tsmap} by the black cross and black circle respectively. Since we know that \sn is a large source, we also measured the extension for an initial disk matching the TeV shape ($R_{\rm init} = 0\fdg25$). In this case, we obtained TS$_{\rm ext} = 7.7$ and $R = 0\fdg26 \pm 0\fdg03_{\rm stat}$ (the green dashed circle in the lower panel of Figure~\ref{fig:tsmap}). Although the extension is still not significantly detected, this result indicates that we are only limited by the statistics and that the extension should be revealed with a few years of additional data. Then we used {\tt gtlike} to perform a spectral analysis between 1 GeV and 2 TeV to evaluate the spectral index. For the point source, the uniform disk and the \hess template, we obtained $\Gamma = 1.55$, $\Gamma = 1.57$ and $\Gamma = 1.79$ respectively, with errors of $\pm 0.17_{\rm stat} \pm 0.27_{\rm syst}$. When taking into account uncertainties, these results are consistent with $1.9 \pm 0.3$, published in \cite{Xing2016}. Spectral results are summarized in Table~\ref{tab:results}.

The four points of the SED were computed with the \hess template, for the reasons described in Section~\ref{sc:hjdt_results} in the case of \hjdt. The resulting plot (lower panel of Figure~\ref{fig:sed}) shows a good overlap between the GeV points and the TeV data. If one compares the spectral points from this work with those published in \cite{Xing2016}, one may observe a disagreement around 5~GeV. But one should note that the SED from the previous work was computed under the assumption of a point source. Our spectral points are also at the level of the upper limits derived by \cite{Acero2015}.

Since \sn is known to be a very symmetric shell-type remnant, the apparent asymmetry in the $\gamma$-ray emission detected by \fermi is quite intriguing. Therefore, we studied the NE and the SW limbs of the remnant separately. To do so, the \hess template was divided into two parts and their spectral parameters fit independently. The division axis was defined as the symmetry axis of the X-ray emission as seen by XMM-Newton  \citep[see Figure~1 in][]{Miceli2012}. The NE region was significantly detected (TS = 28.3, $\Gamma = 1.47 \pm 0.26_{\rm stat}$) but the SW region was not (TS = 12.9, $\Gamma = 2.60 \pm 0.80_{\rm stat}$). Comparing the log-likelihood obtained, we found an improvement of 10.5 when dividing the \hess template, corresponding to a TS value of 21.0 with two additional degrees of freedom ($3.6\sigma$). This is a first indication of an asymmetry of the high-energy $\gamma$-ray emission.

\subsection{Discussion}

The morphology of the emission detected by the LAT is not as puzzling as for \hjdt and its association with \sn is more easily demonstrated. 

Even if the $\gamma$-ray emission detected by \fermi is not found to be significantly extended when fitting a uniform disk, it is consistent with a large spatial model like the \hess template. Because of its high latitude ($14^\circ$ above the Galactic plane) this emission can hardly be due to bad modeling of the Galactic diffuse emission. As for \hjdt, various hypotheses concerning the origin of this signal were considered (e.g., an AGN, a pulsar) but here again none of them were conclusive: no such sources were found at these coordinates and no flux variability was observed. Thus, from the point of view of the position, an association between the GeV signal and \sn is reasonable. When we consider the measured spectrum, any explanation other than the SNR hypothesis seems very unlikely. Indeed, as presented in Section~\ref{sc:sn_results}, the bright peak exhibits a hard spectrum ($\Gamma \sim 1.5$ when fit as a point source and $\sim 1.8$ when fit with the \hess template) corresponding to the typical hard spectral shape of a $\gamma$-ray emission dominated by the inverse Compton component, a scenario observed for several young SNRs, \textit{e.g.}, \rcw and \rxj with a photon index of 1.42 \citep{Ajello2016} and 1.53 \citep{Federici2015} respectively.

If this emission is the GeV counterpart of SN 1006, the LAT data show an asymmetry between the NE and the SW limbs, only the former being detected. The likelihood favors separate fits for the two limbs by $3.6\sigma$ whereas the X-ray and TeV morphologies are rather symmetrical. The insignificant detection of the SW region suggests a fainter IC emission in this part of the remnant than in the NE part, although it is hard to explain since the ambient photon fields are identical and the shock velocity is $\sim 5000$~km$^{-1}$ in both regions \citep{Winkler2014}. But another point should be discussed : the spectral difference between the NE and the SW region while there is not any spectral difference in the TeV energies. This result should be considered in the context of two previous works: \cite{Dubner2002} revealed the presence of a dense \ion{H}{1} cloud  matching the morphology of the southwestern rim of \sn and \cite{Miceli2014} found evidence of interaction between this \ion{H}{1} cloud and the shock of \sn. Thus, the SW region of \sn shows evidence of efficient particle acceleration (magnetic field lines perpendicular to the shock front) as well as an interaction between the shock and a dense medium : $\gamma$-rays of hadronic origin are expected here. For this reason, the somewhat softer spectrum of the faint signal coming from the SW part of the remnant (if confirmed) could be a hint in this direction. \cite{Miceli2016} predicted two hadronic components, one from the shocked cloud and one from the shocked interstellar medium, the latter being too faint to be detectable. The hadronic component related to the cloud is characterized by a low cutoff ($\sim 3$~TeV) and drops drastically above 100~GeV. Thus, the presence of this component should not affect the very-high energy $\gamma$-ray emission which remains dominated by the IC component. This interpretation relies on the assumption of a spectral difference between the NE and the SW region. The large uncertainties on $\Gamma_{\rm SW}$ make this difference not significant and more data are crucially needed to shed light on this point. 
Finally, the presence or not of a hadronic component does not explain why the SW region is fainter than the NE region in the GeV band. This could mean that the IC emission is dimmer in this energy band, for instance due to a harder electron population in the SW region where part of the shock is interacting with a cloud. Again, more data is needed to confirm this effect.

Since the new spectral points for the whole remnant are very close to the upper limits from \cite{Acero2015}, our modeling is consistent with the one shown in the upper panel of Figure~6 in \cite{Miceli2016} ; we confirm that a total hadronic energy of $5 \times 10^{49}$ erg is conceivable to explain the HE $\gamma$-ray emission in the SW region of \sn.

\section{Conclusion}

In this work, we have studied the shell-type SNRs \hjdt and \sn in the GeV band. Analyzing eight years of LAT Pass 8 data in the energy range 1 GeV $-$ 2 TeV, we found significant $\gamma$-ray excesses at the positions of both sources. The spectral study using templates derived from the \hess excess maps revealed a photon index of $\Gamma = 1.66 \pm 0.16_{\rm stat} \pm 0.12_{\rm syst}$ for \hjdt and $\Gamma = 1.79 \pm 0.17_{\rm stat} \pm 0.27_{\rm syst}$ for \sn, both in agreement with previous works \citep{Acero2015, Xing2016}. Although a detection of these two sources with \fermi was not expected without several more years of data, as argued in \cite{Acero2015}'s conclusion, the improved efficiency of the Pass~8 data combined with an energy range extended to higher energy made it feasible.

\begin{figure}[ht]
  \centering
  \includegraphics[width=\columnwidth]{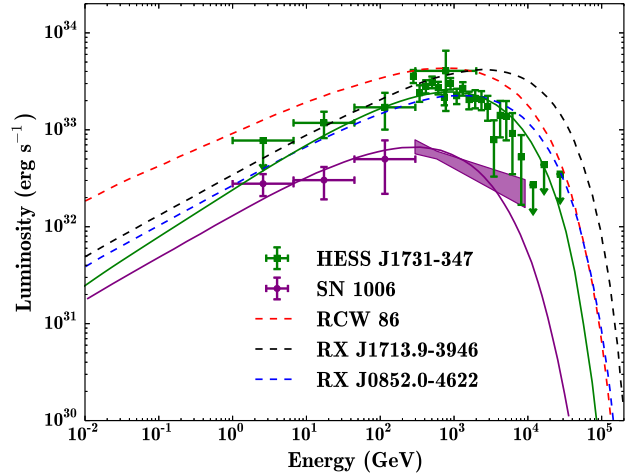}
  \caption{Leptonic modeling for the five TeV shell-type remnants: \rcw (Red), \rxj (Black), \vela (Blue), \hjdt (Green) and \sn (Purple). Spectral points in the GeV and TeV bands are shown for \hjdt and \sn. For the references of each modeling, see \cite{Acero2015} from which the original version of this figure is extracted. \label{fig:snr_modeling}}
\end{figure}

In Figure~\ref{fig:snr_modeling}, we reproduce Figure~3 from \cite{Acero2015} and overlay \hjdt and \sn spectral points in the GeV--TeV band. We can see that the new spectra are in good agreement with the models. Overall, the hard spectra of these SNRs suggest a common scenario in which the bulk of the $\gamma$-ray emission is produced by inverse Compton scattering of high energy electrons. The $\gamma$-ray emission is likely to be leptonic dominated. However, this does not rule out efficient hadron acceleration in these TeV shells and the spectral asymmetry visible in \sn might be a first evidence in this respect. \\

\textit{Acknowledgments}. The \textit{Fermi} LAT Collaboration acknowledges generous ongoing support from a number of agencies and institutes that have supported both the development and the operation of the LAT as well as scientific data analysis. These include the National Aeronautics and Space Administration and the Department of Energy in the United States, the Commissariat \`a l'Energie Atomique and the Centre National de la Recherche Scientifique / Institut National de Physique Nucl\'eaire et de Physique des Particules in France, the Agenzia Spaziale Italiana and the Istituto Nazionale di Fisica Nucleare in Italy, the Ministry of Education, Culture, Sports, Science and Technology (MEXT), High Energy Accelerator Research Organization (KEK) and Japan Aerospace Exploration Agency (JAXA) in Japan, and the K.~A.~Wallenberg Foundation, the Swedish Research Council and the Swedish National Space Board in Sweden. 
 
Additional support for science analysis during the operations phase is gratefully acknowledged from the Istituto Nazionale di Astrofisica in Italy and the Centre National d'\'Etudes Spatiales in France.


\begin{thebibliography}{}
\expandafter\ifx\csname natexlab\endcsname\relax\def\natexlab#1{#1}\fi
\providecommand{\url}[1]{\href{#1}{#1}}

\bibitem[{{Abdo} {et~al.}(2011){Abdo}, {Ackermann}, {Ajello}, {Allafort},
  {Baldini}, {Ballet}, {Barbiellini}, {Baring}, {et~al.}}]{Abdo2011}
{Abdo}, A.~A., {Ackermann}, M., {Ajello}, M., {et~al.} 2011, \apj, 734, 28

\bibitem[{{Acero} {et~al.}(2010){Acero}, {Aharonian}, {Akhperjanian}, {Anton},
  {Barres de Almeida}, {Bazer-Bachi}, {et~al.}}]{Acero2010}
{Acero}, F., {Aharonian}, F., {Akhperjanian}, A.~G., {et~al.} 2010, \aap, 516,
  A62

\bibitem[{{Acero} {et~al.}(2015{\natexlab{a}}){Acero}, {Lemoine-Goumard},
  {Renaud}, {Ballet}, {Hewitt}, {Rousseau}, \& {Tanaka}}]{Acero2015}
{Acero}, F., {Lemoine-Goumard}, M., {Renaud}, M., {et~al.} 2015{\natexlab{a}},
  \aap, 580, A74

\bibitem[{{Acero} {et~al.}(2015{\natexlab{b}}){Acero}, {Ackermann}, {Ajello},
  {Albert}, {Atwood}, {Axelsson}, {Baldini}, {Ballet}, {et~al.}}]{3FGL}
{Acero}, F., {Ackermann}, M., {Ajello}, M., {et~al.} 2015{\natexlab{b}}, \apjs,
  218, 23

\bibitem[{{Acero} {et~al.}(2016{\natexlab{a}}){Acero}, {Ackermann}, {Ajello},
  {Albert}, {Baldini}, {Ballet}, {Barbiellini}, {Bastieri},
  {et~al.}}]{Acero2016a}
---. 2016{\natexlab{a}}, \apjs, 223, 26

\bibitem[{{Acero} {et~al.}(2016{\natexlab{b}}){Acero}, {Ackermann}, {Ajello},
  {Baldini}, {Ballet}, {Barbiellini}, {Bastieri}, {Bellazzini},
  {et~al.}}]{SNRCAT}
---. 2016{\natexlab{b}}, \apjs, 224, 8

\bibitem[{{Ackermann} {et~al.}(2012){Ackermann}, {Ajello}, {Albert},
  {Allafort}, {Atwood}, {Axelsson}, {Baldini}, {Ballet},
  {et~al.}}]{Ackermann2012}
{Ackermann}, M., {Ajello}, M., {Albert}, A., {et~al.} 2012, \apjs, 203, 4

\bibitem[{{Aharonian} {et~al.}(2005{\natexlab{a}}){Aharonian}, {Akhperjanian},
  {Aye}, {Bazer-Bachi}, {Beilicke}, {Benbow}, {et~al.}}]{HGPS2005}
{Aharonian}, F., {Akhperjanian}, A.~G., {Aye}, K.-M., {et~al.}
  2005{\natexlab{a}}, Science, 307, 1938

\bibitem[{{Aharonian} {et~al.}(2008){Aharonian}, {Akhperjanian}, {Barres de
  Almeida}, {Bazer-Bachi}, {Behera}, {Beilicke}, {et~al.}}]{Aharonian2008}
{Aharonian}, F., {Akhperjanian}, A.~G., {Barres de Almeida}, U., {et~al.} 2008,
  \aap, 477, 353

\bibitem[{{Aharonian} {et~al.}(2005{\natexlab{b}}){Aharonian}, {Akhperjanian},
  {Bazer-Bachi}, {Beilicke}, {Benbow}, {Berge}, {Bernl{\"o}hr},
  {et~al.}}]{Aharonian2005}
{Aharonian}, F., {Akhperjanian}, A.~G., {Bazer-Bachi}, A.~R., {et~al.}
  2005{\natexlab{b}}, \aap, 437, L7

\bibitem[{{Aharonian} {et~al.}(2006){Aharonian}, {Akhperjanian}, {Bazer-Bachi},
  {Beilicke}, {Benbow}, {Berge}, {Bernl{\"o}hr}, {et~al.}}]{HGPS2006}
---. 2006, \apj, 636, 777

\bibitem[{{Aharonian} {et~al.}(2009){Aharonian}, {Akhperjanian}, {de Almeida},
  {Bazer-Bachi}, {Behera}, {Beilicke}, {et~al.}}]{Aharonian2009}
{Aharonian}, F., {Akhperjanian}, A.~G., {de Almeida}, U.~B., {et~al.} 2009,
  \apj, 692, 1500

\bibitem[{{Aharonian} {et~al.}(2004){Aharonian}, {Akhperjanian}, {Aye},
  {Bazer-Bachi}, {Beilicke}, {Benbow}, {et~al.}}]{Aharonian2004}
{Aharonian}, F.~A., {Akhperjanian}, A.~G., {Aye}, K.-M., {et~al.} 2004, \nat,
  432, 75

\bibitem[{{Ajello} {et~al.}(2016){Ajello}, {Baldini}, {Barbiellini},
  {Bastieri}, {Bellazzini}, {Bissaldi}, {Bloom}, {Bonino},
  {et~al.}}]{Ajello2016}
{Ajello}, M., {Baldini}, L., {Barbiellini}, G., {et~al.} 2016, \apj, 819, 98

\bibitem[{{Atwood} {et~al.}(2013){Atwood}, {Albert}, {Baldini}, {Tinivella},
  {Bregeon}, {Pesce-Rollins}, {Sgr{\`o}}, {Bruel}, {Charles}, {Drlica-Wagner},
  {Franckowiak}, {Jogler}, {Rochester}, {Usher}, {Wood}, {Cohen-Tanugi}, \&
  {S.~Zimmer for the Fermi-LAT Collaboration}}]{Atwood2013}
{Atwood}, W., {Albert}, A., {Baldini}, L., {et~al.} 2013, 4th Fermi Symposium
  proceedings - eConf C121028, arXiv:1303.3514

\bibitem[{{Atwood} {et~al.}(2009){Atwood}, {Abdo}, {Ackermann}, {Althouse},
  {Anderson}, {Axelsson}, {Baldini}, {Ballet}, {Band}, {Barbiellini}, \&
  et~al.}]{Atwood2009}
{Atwood}, W.~B., {Abdo}, A.~A., {Ackermann}, M., {et~al.} 2009, \apj, 697, 1071

\bibitem[{{Cassam-Chena{\"i}} {et~al.}(2004){Cassam-Chena{\"i}},
  {Decourchelle}, {Ballet}, {Sauvageot}, {Dubner}, \&
  {Giacani}}]{CassamChenai2004}
{Cassam-Chena{\"i}}, G., {Decourchelle}, A., {Ballet}, J., {et~al.} 2004, \aap,
  427, 199

\bibitem[{{Cui} {et~al.}(2016){Cui}, {P{\"u}hlhofer}, \&
  {Santangelo}}]{Cui2016}
{Cui}, Y., {P{\"u}hlhofer}, G., \& {Santangelo}, A. 2016, \aap, 591, A68

\bibitem[{{Doroshenko} {et~al.}(2017){Doroshenko}, {P{\"u}hlhofer}, {Bamba},
  {Acero}, {Tian}, {Klochkov}, \& {Santangelo}}]{Doroshenko2017}
{Doroshenko}, V., {P{\"u}hlhofer}, G., {Bamba}, A., {et~al.} 2017, arXiv:
  1708.04110

\bibitem[{{Dubner} {et~al.}(2002){Dubner}, {Giacani}, {Goss}, {Green}, \&
  {Nyman}}]{Dubner2002}
{Dubner}, G.~M., {Giacani}, E.~B., {Goss}, W.~M., {Green}, A.~J., \& {Nyman},
  L.-{\AA}. 2002, \aap, 387, 1047

\bibitem[{{Enomoto} {et~al.}(2002){Enomoto}, {Tanimori}, {Naito}, {Yoshida},
  {Yanagita}, {Mori}, {Edwards}, {Asahara}, {Bicknell}, {Gunji}, {Hara},
  {Hara}, {Hayashi}, {Itoh}, {Kabuki}, {Kajino}, {Katagiri}, {Kataoka},
  {Kawachi}, {Kifune}, {Kubo}, {Kushida}, {Maeda}, {Maeshiro}, {Matsubara},
  {Mizumoto}, {Moriya}, {Muraishi}, {Muraki}, {Nakase}, {Nishijima}, {Ohishi},
  {Okumura}, {Patterson}, {Sakurazawa}, {Suzuki}, {Swaby}, {Takano}, {Takano},
  {Tokanai}, {Tsuchiya}, {Tsunoo}, {Uruma}, {Watanabe}, \&
  {Yoshikoshi}}]{Enomoto2002}
{Enomoto}, R., {Tanimori}, T., {Naito}, T., {et~al.} 2002, Nature, 416, 823

\bibitem[{{Federici} {et~al.}(2015){Federici}, {Pohl}, {Telezhinsky},
  {Wilhelm}, \& {Dwarkadas}}]{Federici2015}
{Federici}, S., {Pohl}, M., {Telezhinsky}, I., {Wilhelm}, A., \& {Dwarkadas},
  V.~V. 2015, \aap, 577, A12

\bibitem[{{Halpern} \& {Gotthelf}(2010)}]{Halpern2010}
{Halpern}, J.~P., \& {Gotthelf}, E.~V. 2010, \apj, 710, 941

\bibitem[{{Helene}(1983)}]{Helene1983}
{Helene}, O. 1983, Nuclear Instruments and Methods in Physics Research, 212,
  319

\bibitem[{{H.E.S.S.~Collaboration} {et~al.}(2011){H.E.S.S.~Collaboration},
  {Abramowski}, {Acero}, {Aharonian}, {Akhperjanian}, {Anton},
  {et~al.}}]{Acero2011}
{H.E.S.S.~Collaboration}, {Abramowski}, A., {Acero}, F., {et~al.} 2011, \aap,
  531, A81

\bibitem[{{Kargaltsev} {et~al.}(2002){Kargaltsev}, {Pavlov}, {Sanwal}, \&
  {Garmire}}]{Kargaltsev2002}
{Kargaltsev}, O., {Pavlov}, G.~G., {Sanwal}, D., \& {Garmire}, G.~P. 2002,
  \apj, 580, 1060

\bibitem[{{Katagiri} {et~al.}(2005){Katagiri}, {Enomoto}, {Ksenofontov},
  {Mori}, {Adachi}, {Asahara}, {Bicknell}, {et~al.}}]{Katagiri2005}
{Katagiri}, H., {Enomoto}, R., {Ksenofontov}, L.~T., {et~al.} 2005, \apjl, 619,
  L163

\bibitem[{{Kerr}(2011)}]{Kerr2011}
{Kerr}, M. 2011, PhD thesis, University of Washington

\bibitem[{{Klochkov} {et~al.}(2015){Klochkov}, {Suleimanov}, {P{\"u}hlhofer},
  {Yakovlev}, {Santangelo}, \& {Werner}}]{Klochkov2015}
{Klochkov}, D., {Suleimanov}, V., {P{\"u}hlhofer}, G., {et~al.} 2015, \aap,
  573, A53

\bibitem[{{Koyama} {et~al.}(1995){Koyama}, {Petre}, \& {Gotthelf}}]{Koyama1995}
{Koyama}, K., {Petre}, R., \& {Gotthelf}, E.~V., e.~a. 1995, \nat, 378, 255

\bibitem[{{Lande} {et~al.}(2012){Lande}, {Ackermann}, {Allafort}, {Ballet},
  {Bechtol}, {Burnett}, {Cohen-Tanugi}, {Drlica-Wagner}, {Funk}, {Giordano},
  {Grondin}, {Kerr}, \& {Lemoine-Goumard}}]{Lande2012}
{Lande}, J., {Ackermann}, M., {Allafort}, A., {et~al.} 2012, \apj, 756, 5

\bibitem[{{Mattox} {et~al.}(1996){Mattox}, {Bertsch}, {Chiang}, {Dingus},
  {Digel}, {Esposito}, {Fierro}, {Hartman}, {Hunter}, {Kanbach}, {Kniffen},
  {Lin}, {Macomb}, {Mayer-Hasselwander}, {Michelson}, {von Montigny},
  {Mukherjee}, {Nolan}, {Ramanamurthy}, {Schneid}, {Sreekumar}, {Thompson}, \&
  {Willis}}]{Mattox1996}
{Mattox}, J.~R., {Bertsch}, D.~L., {Chiang}, J., {et~al.} 1996, \apj, 461, 396

\bibitem[{{Maxted} {et~al.}(2015){Maxted}, {Rowell}, {de Wilt}, {Burton},
  {Renaud}, {Fukui}, {Hawkes}, {Blackwell}, {Voisin}, {Lowe}, \&
  {Aharonian}}]{Maxted2015}
{Maxted}, N., {Rowell}, G., {de Wilt}, P., {et~al.} 2015, arXiv:1503.06717

\bibitem[{{McClure-Griffiths} {et~al.}(2005){McClure-Griffiths}, {Dickey},
  {Gaensler}, {Green}, {Haverkorn}, \& {Strasser}}]{SGPS2005}
{McClure-Griffiths}, N.~M., {Dickey}, J.~M., {Gaensler}, B.~M., {et~al.} 2005,
  \apjs, 158, 178

\bibitem[{{Miceli} {et~al.}(2014){Miceli}, {Acero}, {Dubner}, {Decourchelle},
  {Orlando}, \& {Bocchino}}]{Miceli2014}
{Miceli}, M., {Acero}, F., {Dubner}, G., {et~al.} 2014, \apjl, 782, L33

\bibitem[{{Miceli} {et~al.}(2012){Miceli}, {Bocchino}, {Decourchelle},
  {Maurin}, {Vink}, {Orlando}, {Reale}, \& {Broersen}}]{Miceli2012}
{Miceli}, M., {Bocchino}, F., {Decourchelle}, A., {et~al.} 2012, \aap, 546, A66

\bibitem[{{Miceli} {et~al.}(2016){Miceli}, {Orlando}, {Pereira}, {Acero},
  {Katsuda}, {Decourchelle}, {Winkler}, {Bonito}, {Reale}, {Peres}, {Li}, \&
  {Dubner}}]{Miceli2016}
{Miceli}, M., {Orlando}, S., {Pereira}, V., {et~al.} 2016, \aap, 593, A26

\bibitem[{{Muraishi} {et~al.}(2000){Muraishi}, {Tanimori}, {Yanagita},
  {Yoshida}, {Moriya}, {Kifune}, {Dazeley}, {Edwards}, {Gunji}, {Hara}, {Hara},
  {Kawachi}, {Kubo}, {Matsubara}, {Mizumoto}, {Mori}, {Muraki}, {Naito},
  {Nishijima}, {Patterson}, {Rowell}, {Sako}, {Sakurazawa}, {Susukita},
  {Tamura}, \& {Yoshikoshi}}]{Muraishi2000}
{Muraishi}, H., {Tanimori}, T., {Yanagita}, S., {et~al.} 2000, \aap, 354, L57

\bibitem[{{Tanaka} {et~al.}(2011){Tanaka}, {Allafort}, {Ballet}, {Funk},
  {Giordano}, {Hewitt}, {Lemoine-Goumard}, {Tajima}, {Tibolla}, \&
  {Uchiyama}}]{Tanaka2011}
{Tanaka}, T., {Allafort}, A., {Ballet}, J., {et~al.} 2011, \apjl, 740, L51

\bibitem[{{The Fermi-LAT Collaboration}(2017)}]{3FHL}
{The Fermi-LAT Collaboration}. 2017, arXiv:1702.00664

\bibitem[{{Tian} {et~al.}(2008){Tian}, {Leahy}, {Haverkorn}, \&
  {Jiang}}]{Tian2008}
{Tian}, W.~W., {Leahy}, D.~A., {Haverkorn}, M., \& {Jiang}, B. 2008, \apjl,
  679, L85

\bibitem[{{Tian} {et~al.}(2010){Tian}, {Li}, {Leahy}, {Yang}, {Yang},
  {Yamazaki}, \& {Lu}}]{Tian2010}
{Tian}, W.~W., {Li}, Z., {Leahy}, D.~A., {et~al.} 2010, \apj, 712, 790

\bibitem[{{Winkler} {et~al.}(2014){Winkler}, {Williams}, {Reynolds}, {Petre},
  {Long}, {Katsuda}, \& {Hwang}}]{Winkler2014}
{Winkler}, P.~F., {Williams}, B.~J., {Reynolds}, S.~P., {et~al.} 2014, \apj,
  781, 65

\bibitem[{{Xing} {et~al.}(2016){Xing}, {Wang}, {Zhang}, \& {Chen}}]{Xing2016}
{Xing}, Y., {Wang}, Z., {Zhang}, X., \& {Chen}, Y. 2016, \apj, 823, 44

\bibitem[{{Yuan} {et~al.}(2014){Yuan}, {Huang}, {Liu}, \& {Zhang}}]{Yuan2014}
{Yuan}, Q., {Huang}, X., {Liu}, S., \& {Zhang}, B. 2014, \apjl, 785, L22

\end{thebibliography}

\end{document}